\documentclass[11pt]{article}
\usepackage[utf8]{inputenc}  % gebruik de juiste 'character encoding'
\usepackage{hyperref}        % geef URLs netjes weer
\usepackage{booktabs}		 % mooiere tabellen
\usepackage{a4wide}          % papierformaat en marges
\title{Accelerating Molecular H$_2^+$ Beam in AGS and RHIC}
\author{Xiaodong Jiang, Brookhaven National Laboratory}
\date{Technical Report (BNL-229345-2026-TECH), January 12, 2026} 

\begin{document}
	
	\maketitle                  % maakt een kopje met titel, auteur en datum
	\thispagestyle{empty}       % zet paginanummering uit voor deze pagina
	
	% kopje met bijbehorende titel; varianten: \subsection, \subsubsection
	\section*{Summary:}
	
	Before RHIC shutdown, it is urgent to determine if molecular H$_2^+$  beam can be accelerated in RHIC up to 100 GeV/u. 
	After the recent success of acceleration molecular H$_2^+$ beam (from EBIS) in Booster to 1.0 GeV/u kinetic energy,  we propose to carry out tests to accelerate H$_2^+$ beam in AGS (up to 12 GeV/u) and to RHIC (up to 100 GeV/u).  In this Note, we address the main effects that might hamper H$_2^+$ beam acceleration, outline the impacts and benefits of high energy H$_2^+$ beam to BNL.   	
	
\section*{Effects that Limit H$_2^+$ Beam Acceleration}	
Recently,in preparation for NSRL's 2025 NASA biology run,  H$_2^+$ ions injected from EBIS was  accelerated to 1.0 GeV/u kinetic energy in Booster, with an intensity of $1.4 \times10^{10}$ proton/spill when extracted to Booster R-line (NSRL), provided a relative lower cost option of proton beams compared to that of Tandem and Linac.
	
 The acceleration of H$_2^+$ beam has been studied to provide proton beams at high flux, for example in cyclotron accelerators\cite{Winklehner:2017kcf, Koay_2023, Winklehner_2023}.  Aside from limited H$_2^+$ flux from injector, the main beam intensity limitations come from two effects: i, Lorentz Effect: static electrical force that tears apart H$_2^+$, caused by high E field (in rest-frame) originated from high B field (in lab-frame)  ii, dissociation of H$_2^+$ molecules while colliding with beam gas.
 
 \subsection*{Lorentz Effect: strong electrical field in rest-frame that could tear H$_2^+$ apart}
 When H$_2^+$ experience a bending magnetic field $\vec{\textbf{B}}$ perpendicular to its velocity ($\vec{\textbf{v}}$)in lab-frame, an electrical field $\vec{\textbf{E}}^{\prime}$ is experienced in its rest-frame, along $\vec{\textbf{v}} \times \vec{\textbf{B}}$, with field strength  $E^{\prime}_x = \gamma \beta c B_y$, where $\gamma$ and $\beta$ are the frame boost factors. As a semi-classical estimation, when  $E^{\prime}_x$ field acting upon electron's obit ($a_0$ $\approx$ Bohr radius 0.53 $\AA$) produces a ''Lorentz-separation-potential'' ($E^{\prime} \cdot a_0$) comparable to H$_2^+$ dissociation energy, the chance of H$_2^+$ at high-lying vibration states to become unstable increase and a proton to “leak"
 from the potential well will lead to beam loss.   Table-1 lists H$_2^+$ beam energy, dipole bending field $B_y$ for H$_2^+$ beam in the ring, and Lorentz energy $E^{\prime}_x \cdot a_0$ for Booster (at 1 GeV/u kinetic energy), AGS (at 12 GeV/c/u momentum), and RHIC (at 20, 50, 100 and 137.5 GeV/c/u momentum). In comparison, the dissociation energy needed to breakup H$_2^+$ molecule is 2.65 eV.

\begin{table}[h]
	\centering
	\caption{H$_2^+$ beam energy, typical dipole bending magnetic field $B_y$ in lab-frame, and $E^{\prime}_x$ in rest-frame, and estimated ''Lorentz-separation-potential'' $E^{\prime}_x \cdot a_0$}
	\begin{tabular}{llccc}
		\toprule
		\cmidrule(lr){2-5}
		  & Energy/mom.  & $B_y$ (Lab) & $E^{\prime}_x$ (rest-frame) & Lorentz $E^{\prime}_x \cdot a_0$ \\
		  & (GeV/u, GeV/c/u)   & (Tesla) & (10$^{9}$ Volt/meter) & (eV) \\
		\midrule
		Booster  & 1.0 (kinetic energy) & 0.352 & 0.19 &  0.01 \\
		 AGS  & 12.0 (momentum)& 0.982 &3.77  &  0.20 \\
		RHIC  & 20.0 & 0.551 & 3.52 &  0.19 \\
		      & 50.0 & 1.377 & 22.0 & 1.17 \\
		      & 100.0 & 2.754 & 88.0  & 4.67 \\
		EIC (max.)  & 137.5 & 3.786 & 166.0 &  8.82 \\
		\bottomrule
    \end{tabular}
\end{table}

Using the semi-classical estimation as a guidance, we conclude that accelerating H$_2^+$ in AGS to 12 GeV/u will not likely to be prohibited by Lorentz effect. In the RHIC Ring at 50 $\sim$ 100 GeV/u (and at EIC max. 137.5 GeV/u), while approaching the ''strong field atomic regime'', an upper limit of  H$_2^+$ acceleration ''might be encountered'' due to Lorentz effect.  Therefore, before RHIC shutdown, it is urgent to determine whether such an energy upper limit exists for H$_2^+$ acceleration in RHIC such that operation options can be considered for decades to come.
 
 \subsection*{Dissociation Effect: beam gas collisions that rip H$_2^+$ molecule apart}
 
As a reference point, H$^-$ ion (electron binding energy 0.75 eV) stripping to H$^0$ cross section at 800 MeV is $8\times10^{-19}$ cm$^2$ \cite{PhysRevA.53.3201}. It was further confirmed
at J-PARC\cite{OKABE201611} and accepted at FNAL\cite{Neuffer_2023} that H$^-$ dissociation cross-section at 1 GeV is: $\sigma \sim 5 \times10^{-19}$ cm$^2$. Furthermore, at higher energies, the dissociation cross section is not expected to change significantly (Bethe-Born scaling).
Although it was measured precisely, we can conservatively assume that H$_2^+$ (binding energy 2.67 eV) dissociation cross section is $5 \times10^{-19}$ cm$^2$ for energy between 1 GeV/u ~100 GeV/u. 
 
Table~2 lists typical vacuum levels, beam gas number density N$_a$, dissociation collision mean-free-path $\lambda = 1/N_a \sigma$  for Booster, AGS and RHIC Ring.  In  the worst vacuum case of AGS, as of Jan. 5th, 2026, most AGS vacuum gauges were at $<$ 2E-9 torr level (except for a03 and e17: 1.1E-8 torr), the collision mean-free-path is at least 5-orders of magnitude higher than the ring size. Notice that the vacuum level assumption in AGS is somewhat different compared to the worst case assumption used in an earlier study \cite{Fischer2008Injection}. Under the very conservative assumption of the dissociation cross section, the  expected beam life time in RHIC ($\tau \approx \lambda/\beta c$) is at the level of $>3$ minutes. 
       
\begin{table}[h]
	\centering
	\caption{Typical vacuum, beam gas number density N$_a$, collision mean-free-path $\lambda$ to be compared with  ring circumference L,  for Booster, AGS and RHIC ring}
	
	\begin{tabular}{llcccc}
		\toprule
	%	\cmidrule(lr){2-6}
		& Energy & Typical Vacuum  &  N$_a$ & Collision Free & Circum.   \\
		& GeV/u& (torr)   & (cm$^{-3}$) & Path $\lambda$ (km) & L (km) \\
		\midrule
		Booster & 1.0 (kinetic) & 1E-10  & 3.54E6 & 5.66E6 & 0.202  \\
		AGS  & 12 & 2E-9  & 7.07E7  & 2.83E5 & 0.807 \\
		RHIC  & 20-100 & 1E-10 (warm sec.) & 3.54E6 & 5.66E6 & 3.830  \\
		      &        & 1E-11 (cold sec.) & 3.54E5 & 5.66E7 &   \\
		\bottomrule
		\end{tabular}
\end{table}

	\section*{Goal of H$_2^+$ Beam Tests in AGS and RHIC Ring}

The goal of the test is to demonstrate molecular H$_2^+$ beam acceleration in AGS (up to 12 GeV/u), in RHIC ring (Blue ring, up to 100 GeV/u), determine the  maximum energy H$_2^+$ beam can be accelerated.  Quantify beam loss at each stages as references for future improvements. Other technical issues could also be encountered during the tests.

	\section*{Impacts and Benefits to BNL}
	By enhancing BNL's machine capability, the successful acceleration of H$_2^+$ beam in AGS and in RHIC ring will be marked as a significant progress in accelerator physics at BNL. Since H$_2^+$ beam produces only half space charge compared to proton, it opens up many other accelerator technology and facility options for future high intensity proton beams, in applications such as proton therapy, spallation neutron sources, isotope production and in secondary beams and neutrino productions, especially promising in linear accelerators that can avoid Lorentz effects.  The following sections list the direct impact and benefit to the existing and planned BNL operation.

\subsection{Provide an alternative option of proton beam at a lower cost}
As already demonstrated in the Booster R-line (NSRL), H$_2^+$ beam, with electron stripped off while passing through a thin foil, provided proton beams at a reasonable flux  (intensity $\sim$1E10 protons/spill). At a relatively lower cost, H$_2^+$ beam from  EBIS injector offer operation flexibility, reliability and affordability. This practical option also extends to possible future AGS extracted beam facilities, such as HEET and the U-line test beam, as well as EIC.

\subsection{Provide a combination of proton and electron beams in RHIC Ring for EIC}
If successfully accelerated in the RHIC Ring,
H$_2^+$ ions, with two protons and one electron at 2:1 ratio and at the same velocity, will be provided in every beam bunch and in every collision crossing.

Table-3 lists M{\o}ller scattering $e^- + e^- \rightarrow e^- + e^- $ kinematics for EIC with electron beam on electron of molecular H$_2+$ beam.  Electron energy E$_{eH}$ of H$_2^+$ beam,  M{\o}ller pair center-of-mass energy $\sqrt{s}$, electron opening angle and energy in lab corresponding to 90$^\circ$ center-of-mass scattering. 

\begin{table}[h]
	\centering
	\caption{Electron-electron collision kinematics with H$_2^+$ beam in EIC. Electron beam energy (GeV), H$2^+$ beam energy (GeV/u), H$_2^+$ beam momentum, and the corresponding electron's energy E$_{eH}$ in H$_2^+$, center-of-mass M{\o}ller collision energy $\sqrt{s}$, typical angle and energy of M{\o}ller pair in lab-frame corresponding to 90$^\circ$ center-of-mass scattering}
	\begin{tabular}{llcccc}
		\toprule
    	\cmidrule(lr){2-6}
		e-beam& H$_2^+$ mom.  & E$_{eH}$ & $\sqrt{s}$ & $\theta_{1e}=\theta_{2e}$ & E$_{1e}=E_{2e}$  \\
		GeV & GeV/c/u   & GeV & GeV & deg. & GeV \\
		\midrule
		5.0  & 20.0 & 0.0109 & 0.4670 &  5.3475 & 2.5055  \\
	       	 & 50.0 & 0.0272 & 0.7380 & 8.4421 & 2.5136 \\
		     & 100.0 & 0.0545 & 1.0437 & 11.9167 & 2.5272  \\
		     & 137.5 & 0.0749 &   1.2238 & 13.9546 & 2.5374 \\
		\midrule
        12.0 & 20.0 & 0.0109 & 0.7235 &  3.4533 & 6.0055  \\
             & 50.0 &  0.0272 & 1.1434 &  5.4551 & 6.0136 \\
             & 100.0 & 0.0545 & 1.6169 &  7.7084 & 6.0272  \\
             & 137.5 & 0.0749 & 1.8959 &  9.0337 & 6.0374 \\
		\midrule
        18.0 & 20.0  & 0.0109 & 0.8661 &  2.8199 &  9.0055 \\
             & 50.0  & 0.0272 & 1.4003 &  4.4552 &  9.0136 \\
             & 100.0 & 0.0545 & 1.9803 &  6.2970 &  9.0272 \\
             & 137.5 & 0.0749 & 2.3220 &  7.3810 &  9.0374 \\
		\bottomrule
	\end{tabular}	
\end{table}

\subsection{Potential impacts to Electron-Ion-Collider}
Once H$_2^+$ beam acceleration is confirmed in RHIC Ring,
 it could provide an alternative ion for EIC collision which could be especially valuable for the initial instrument commissioning. 

Since all survival H$_2^+$ ions must carry protons and electron at 2:1 ratio at the collision point, $e^- + p$ and $e^- + e^-$ collision luminosity ratio is kept precisely at 2:1 for every bunch crossing. By tagging M{\o}ller scattering $e^- + e^- \rightarrow e^- + e^- $  pair at forward angles, $e^- + p$ collision luminosity become precisely known. Therefore, all $e^- + p$  cross section measurements are effectively anchored to $e^- + e^-$ cross sections which are well-known in QED.

Furthermore, $e^- + e^- \rightarrow e^- + e^- $ events provided by H$_2^+$ beam can be used as a standard tool in detector calibration and shakedowns.  M{\o}ller scattered electron pair have only one kinematic degree-of-freedom, when one electron is tagged by its direction or energy, the other electron's direction and energy is well-determined. The over-constrained kinematic correlations provide ideal benchmark tools to calibrate tracking detector performance as well as electron PID detectors. 
  
More than one decade before EIC commissioning, it is urgent to know the full RHIC machine capacity in terms of ion species available such that detector design and commissioning can be planned accordingly.

\section*{Future Improvement Options}
	
	Further studies and tests during RHIC shutdown should also be carried out with  other molecular beams  such as $H_3^+$ and $D_2^+$, and non-fully-stripped ions, such as $^3$He$^+$ (electron binding energy 54.4 eV), injected from EBIS to Booster and to AGS.

	\section*{Appendix: Properties of H$_2^+$ Molecule}
	Molecular hydrogen ion H$_2^+$, one electron shared by two protons, has charge: Q=1.\\
	Mass = 2$m_p$+$m_e$ = 1877.055 MeV \\
	Equilibrium bond distance between two protons $\approx 1.06 \AA$
	Size of electron cloud, comparable to atomic hydrogen:  $\left\langle r \right\rangle  \approx 0.53 \AA$ \\
	Energy needed to break H$_2^+$ molecule:  2.65 eV, i.e. $\mathrm{H_2^+ \rightarrow}  \mathrm{H(1s) + p}$ \\
	Dissociation cross section at 1 GeV/u (kinetic), 12 GeV/u and 100 GeV/u, $< 5\times 10^{-19}$ cm$^2$\\

\bibliographystyle{unsrt}
\bibliography{references}
\end{document}